\documentclass[onecolumn]{emulateapj}
\usepackage{graphicx}
\def\apj{{\it ApJ \,}}

\def\ea{\ et al. \,}

\def\be{\begin{equation}}
\def\ee{\end{equation}}

\shorttitle{Steady-State Synchrotron Spectra}
\shortauthors{Rephaeli \& Persic}
\begin{document}

%\title{Synchrotron Spectrum from a Steady-State Electron Distribution}
\title{Synchrotron and Compton Spectra from a Steady-State Electron Distribution}

\author{Y. Rephaeli\altaffilmark{1}}
\affil{School of Physics and Astronomy, Tel Aviv University, Tel Aviv, 69978, Israel}
\email{yoelr@wise.tau.ac.il}

\and

%\author{M. Persic}
\author{M. Persic\altaffilmark{2}}
\affil{INAF/Osservatorio Astronomico di Trieste and INFN-Trieste, via G.B.Tiepolo 
11, I-34143 Trieste, Italy}

\altaffiltext{1}{Center for Astrophysics and Space Sciences, University of California,
San Diego, La Jolla, CA 92093-0424}

\altaffiltext{2}{INFN-Trieste, via A.Valerio 2, I-34127 Trieste, Italy}

\begin{abstract}
Energy densities of relativistic electrons and protons in extended galactic and intracluster 
regions are commonly determined from spectral radio and (rarely) $\gamma$-ray 
measurements. The time-independent particle spectral density distributions 
are commonly assumed to have a power-law (PL) form over the relevant energy range. A 
theoretical relation between energy densities of electrons and protons is usually adopted, and 
energy equipartition is invoked to determine the mean magnetic field strength in the emitting 
region. We show that for typical conditions, in both star-forming and starburst galaxies, these 
estimates need to be scaled down substantially due to significant energy losses that (effectively) 
flatten the electron spectral density distribution, resulting in a much lower energy density than 
deduced when the distribution is assumed to have a PL form. The steady-state electron distribution 
in the nuclear regions of starburst galaxies is calculated by accounting for Coulomb, bremsstrahlung, 
Compton, and synchrotron losses; the corresponding emission spectra of the latter two processes are 
calculated and compared to the respective PL spectra. We also determine the proton steady-state 
distribution by taking into account Coulomb and $\pi$ production losses, and briefly discuss 
implications of our steady-state particle spectra for estimates of proton energy densities and 
magnetic fields.

\end{abstract}


\section{Introduction}

Nonthermal phenomena in galaxies and galaxy clusters are important for a more complete 
understanding of these systems, and for knowledge of the origin of relativistic particles and 
magnetic fields. Current and future measurement capabilities necessitate fairly detailed 
modeling of these quantities, and inclusion of the impact of interactions of relativistic particles 
in neutral and ionized magnetized media.

Measurements of radiative yields of relativistic electrons and protons across a wide spectral 
range provide the observational basis for determining their spectral density distributions 
(e.g., Schlickeiser 2002). Basic properties of these particles and their broad energy ranges are 
determined from measurements of radio, X-ray, and $\gamma$-ray emission from primary 
electrons and secondary electrons and positrons (produced in $\pi^{\pm}$ decay) in synchrotron, 
bremsstrahlung, and Compton processes, as well as $\gamma$-ray emission from protons 
via $\pi^{0}$ decay.  These spectral bands do not fully cover the particle very broad energy 
ranges, particularly at low 
energies. Consequently, there is an appreciable degree of indeterminacy in quantitative 
estimations of integrated quantities, most important of which are particle energy densities. 
Due to the approximate decreasing (with energy) PL form of the spectral density, 
the generally unknown value of the low energy cutoff introduces substantial uncertainty in 
both proton and electron energy densities. Yet, as we discuss in this paper, the implications of 
this uncertainty are quite significant.

Relativistic particles are one of the consequences of the formation and evolution 
of high-mass stars, so phenomena related to these particles are of interest in all 
star-forming galaxies. Reliable estimates of particle energy densities are very 
much needed for a quantitative evaluation of acceleration models and comparison 
with other measures of stellar evolutionary processes that give rise to 
high-energy phenomena, such as SN explosions, compact remnants of core collapse, 
and SN shocks, all of which are largely gauged by formation rates of high-mass stars 
(Torres \ea 2012, Persic \& Rephaeli 2010).

For the purpose of assessing the significance of realistic estimation of their particle 
energy densities, galaxies in which high-energy phenomena are dominated by stellar 
activity are of particular interest. Among these, well-observed nearby galaxies are 
especially suited, such as the two nearby starburst (SB)galaxies M82 and NGC253, 
for which the gas properties in the nuclear SB region are well determined and their radio 
emission well mapped. Moreover, $\gamma$-ray emission from these galaxies has been 
detected by the {\it Fermi} Large Area Telescope at GeV energies (Ackermann et al. 2012), 
and at TeV energies by the Cherenkov arrays VERITAS and H.E.S.S (Acciari \ea 2009, Acero 
\ea 2009), at flux levels that are in agreement with theoretical predictions (Domingo-Santamaria 
\& Torres 2005, Persic, Rephaeli, \& Arieli 2008, de Cea \ea 2009, Rephaeli \ea 2010). 

In this paper we assess the reliability of estimation of electron energy densities from 
measurements of synchrotron radio emission, quantifying the significant over-estimation of 
particle energy densities in SBGs due to the common assumption of a PL form for the spectral 
density of the emitting electrons. In Section 2 we show that when all relevant electron energy 
losses are accounted for, the electron spectral density can obviously deviate significantly from a 
PL form, as had been quantitatively demonstrated long ago (e.g, Rephaeli 1979, Pohl 1993). 
This necessitates re-calculation of the synchrotron and Compton emissivities (Section 3). As we 
demonstrate in the latter section, normalization of the electron spectrum by comparison 
with radio measurements has significant consequences also for the predicted Compton X-ray 
and $\gamma$-ray spectra. The revised electron spectra yield significantly different estimates of the 
electron energy density than those obtained when the standard formula is used, as 
shown specifically (Section 4) for conditions in SB nuclei. We briefly discuss our results in Section 5.

\section{Steady-State Electron Distribution}

It is commonly assumed that steady-state energetic particle (mostly electrons, protons, 
and helium nuclei) distributions can be well approximated by a PL form for a 
range of values $[ \gamma_{1}, \gamma_{2}]$ of the Lorentz factor, $\gamma$; for 
electrons, the  spectral density is 
\be
N_{pl}(\gamma) = N_{1} \gamma^{-q_{pl}} \, ,
\label{eq:ss_n_e_p}
\ee 
where $N_{1}$ is a normalization constant and $q_{pl}$ is the spectral index. 
Since this commonly used form has no temporal dependence, it is implicitly 
{\it assumed} to be a steady-state distribution. This is the form used in the 
derivation of the standard formulae for  electron synchrotron, bremsstrahlung, 
and Compton spectra. While this approximate single-index PL form may 
be sufficiently accurate for some purposes, it could be very inadequate when the 
relevant electron energy loss processes have different energy dependence.

In a SB nuclear (SBN) region intense star formation yields a high SN rate and consequently 
efficient particle acceleration. A galactic dynamical process, the SB phase is long, O($10^8$) 
year, and since energetic electrons and protons sustain significant energy losses in the relatively 
small gas-rich, magnetically and radiatively intense SBN source region, steady-state is expected. 
The theoretically predicted single index PL spectral density in the acceleration region 
evolves to become a curved steady-state profile as a result of energy loss processes. To 
determine the steady-state electron spectral density, we assume that the {\it initial} spectrum 
of primary accelerated electrons is a PL in momentum. Since we are mostly interested 
in the radiative yields of relativistic electrons with $\gamma \gg 1$, we express the single-index 
spectrum in terms of $\gamma$, and write for the injection rate per unit volume 
\be
\biggl(\frac{dN}{dt}\biggr)_i=k_{i} \gamma^{ - q_{i}} \, ,
\ee
where $k_{i}$ is a normalization constant. When an exact description of the electron spectrum 
is needed (also) at low energies,  $\gamma \sim 1$, the exact relation between energy and 
momentum has to be used, for which the $\gamma$ dependence of the injection spectrum is 
$\gamma (\gamma^{2}-1)^{-(q_{i}+1)/2}$. Whereas the dependence of the electron density 
on $\gamma_1$ can be appreciable (as discussed below), and since typically 
$\gamma_{2} >10^6$, the exact value of this upper cutoff is of little significance for 
relevant values of $q_{i}$.

Energetic particle propagation out of their source region and throughout interstellar (IS) 
space is typically described as a combination of convection and diffusion. We assume that 
in the relatively small SBN region spatial density gradients are not significant, so that 
the particle density is roughly uniform. While particle escape out of the SBN is essentially 
a catastrophic loss that would generally be included as a sink term in the kinetic equation 
describing the spectral distribution, when this term is energy independent its overall 
impact essentially amounts to an overall constant factor in the expression for the (steady 
state) spectral density. Since in our treatment here the overall normalization of the 
electron density is set by the measured level of radio emission, this escape term does 
not have to be explicitly included in the equation for the steady state distribution. 
For our purposes here we therefore focus only on the spectral dependence of the 
steady-state electron density, $N(\gamma)$, which is determined by the energy 
dependence of the total energy loss rate, $-d\gamma/dt =b(\gamma)$. In this case 
the electron steady-state spectral density is determined by solving the simple kinetic equation 
\be
- \frac{d[b(\gamma)N(\gamma)]}{d\gamma} = k_{i}\gamma^{-q_{i}}\, .
\ee

The relevant energy loss processes in a magnetized (H-He) gas permeated by intense (IR) 
radiation field are ionization, electronic excitation 
(or Coulomb), bremsstrahlung, and synchrotron-Compton. The corresponding energy loss rates 
for these well-known processes are $b_{0}(\gamma)$, $b_{1}(\gamma)$, and 
$b_{2}(\gamma)$, respectively. 
In a medium which consists of ionized, neutral, and molecular gas with densities 
$n_i$, $n_H$, and $n_{H2}$, respectively, the lower order energy loss of (a charged energetic 
particle) is by exciting plasma oscillations and by ionization. The expressions 
for the exact loss rate for electron energies down to the sub-relativistic regime 
were calculated by Gould (1972, 1975; see also Schlickeiser 2002); these yield the 
approximate total rate  
\begin{eqnarray}
b_{0}(\gamma) ~\simeq~ \frac {1.1\times 10^{-12} } {\beta}\, \biggl[ n_i \biggl( 1.0 - \frac 
{ln\, {\rm n_i}} {74.6}\biggr) + 0.4 (n_H +2n_{H2}) \biggr] \,\, s^{-1}\, ,
\label{eq:coul_loss}
\end{eqnarray}
where $\beta=(1-\gamma^{-2})^{-1/2}$. 
In the strong shielding limit, an approximate expression for the closely related 
bremsstrahlung loss rate is (Gould 1975) 
\begin{eqnarray}
b_{1}(\gamma)~\simeq~ 1.8\times 10^{-16}~ \gamma ~ [n_{i} +4.5 (n_H + 2n_{H2})] \,\, {\rm s}^{-1}\,. 
\label{eq:bremss_loss}
\end{eqnarray}
The synchrotron-Compton loss rate of an electron in a magnetic field $B$ and in a 
radiation field with energy density $\rho_{\rm r}$ is (e.g., Blumenthal \& Gould 1970) 
\begin{eqnarray}
b_{2}(\gamma) ~ \simeq 
~ 1.3\times 10^{-9} \, \gamma^2\, \bigl( B^2 + 8\pi \rho_{\rm r} \bigr) ~ \, {\rm s}^{-1}\, .
\label{eq:syn_loss}
\end{eqnarray}
We note that the range of electron energies relevant for our discussion does not 
extend to energies beyond the validity of the Thomson limit, i.e., the incident 
photon energy, $\epsilon_0$, in the electron frame, $\gamma \epsilon_0$, satisfies 
the inequality $\gamma \epsilon_0 \ll mc^2$ for all values of $\epsilon_0 <0.01$ 
eV and $\gamma <10^6$ of interest to us here. Obviously, the full Klein-Nishina cross 
section has to be used at much higher energies (e.g., Schlickeiser \& Ruppel 2010). 
Aside from the weak (logarithmic) $\gamma$ dependence in the expressions for $b_0$ 
and $b_1$, the value of the subscript in each of the three loss rates corresponds to the 
power of its $\gamma$ dependence, with the relative magnitude of each term determined 
by the gas densities, $B$, and the radiation field. At low energies the Coulomb rate 
dominates, whereas at high energies synchrotron-Compton losses dominate. 
The steady-state spectral electron density is then
\be
N(\gamma) = \frac{ k_{i}\gamma^{-(q_{i} -1)}}{b(\gamma) (q_{i}-1) }\, . 
\label{eq:n_e_ss}
\ee
Clearly, the different energy dependence of the loss processes results in a curved 
steady-state spectral density, with the value of the effective PL index changing 
from $q_{i}-1$ at energies for which Coulomb losses begin to dominate over the 
combined losses by bremsstrahlung and synchrotron-Compton, at energies well below 
\be
\gamma \leq 3.0\times 10^3  \bigl ( \frac {n_{i}}{100 \, cm^{-3}}\bigr)^{1/2} 
\bigl ( \frac {B}{100 \, \mu G} \bigr)^{-1} ,
\ee
to $q_{i}+1$ at higher energies, for which synchrotron-Compton losses dominate 
over the combined Coulomb and bremsstrahlung losses. 
To quantitatively compare the two distributions, and in order to assess the physical 
implications of using a PL form for the electron spectral density instead of 
the more realistic steady-state form, we first need to set the relative normalization.  
A physically meaningful way of doing so is by normalizing to the same energy density,  
$\rho = \rho_{pl}$, 
\be
\frac {k_{i}}{q_{i}-1} \int_{\gamma_1}^{\gamma_2} \frac {\gamma^{-(q_{i} 
-2 )} d\gamma}{b(\gamma)} = N_{1} \int_{\gamma_1}^{\gamma_2} \gamma^
{-(q_{pl} -1)} d\gamma \, .
\ee
Since the electron density is usually deduced from the synchrotron emission, 
a more observationally based choice is normalization to the measured radio flux. In the 
next section we compare results obtained based on these two different normalizations.

\begin{figure}[h]
\epsscale{0.7}
%\plotone{fig1_den_sbn_2.3.eps}
\plotone{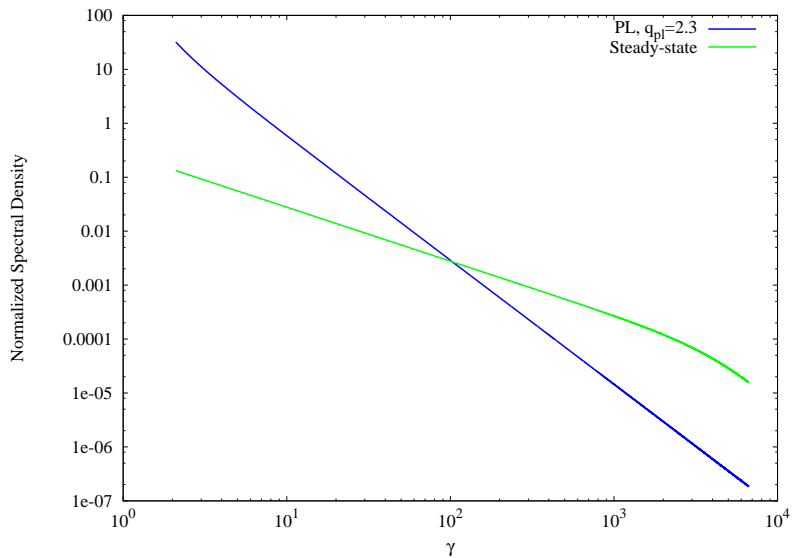}
\caption{
Electron steady-state and PL spectral densities for typical conditions 
in a SBN, normalized to the same energy density. Spectral densities were calculated with SBN 
parameters $q_{pl}=2.3$, $q_{i}=2.0$, $B=100 \, \mu$G, $\rho \simeq 10\, eV/cm^{-3}$, 
$n_i =200$ cm$^{-3}$, $n_H=30\, cm^{-3}$, and $n_{H2}=150\, cm^{-3}$.
\label{fig1}}
\end{figure}

In our numerical estimates we take the theoretically motivated (and observationally supported) 
value of the injection PL index $q_{i}=2$, and the observationally deduced value 
of $q_{pl}$. An estimate of the latter quantity is obtained from the measured PL index 
of the radio spectrum, typically in the range $0.7\pm 0.05$ in the central SB region (as 
in the nearby SBGs M82 \& NGC253; for references to the radio measurements, see Rephaeli, Arieli, 
\& Persic 2010, and Persic \& Rephaeli 2014). Due to the significant (nonlinear) dependence of 
the electron energy density, $\rho_e$, on the spectral index, we adopt the lower ($1\sigma$) 
end of the measured radio index of $0.65$, which corresponds to the fiducial value 
$q_{pl}=2.3$. Doing so we (conservatively) underestimate the impact on estimation of 
$\rho_e$ due to approximating a steady-state spectral density with a PL distribution. 
The PL and steady-state spectral densities are shown in Figure 1 for typical conditions in 
the dense and intense SBN environment, with $B=100\, \mu$G, a total 
radiation field energy density which is estimated to be about $10\, eV/cm^{-3}$ (e.g., 
Porter \& Strong 2005), or $\rho \simeq 40\rho_0$, where $\rho_0\simeq 4.2\times 10^{-13}\, 
erg/cm^{-3}$ is the (present value of the) CMB energy density. Assumed ionized and neutral 
gas densities are $n_i=200\, cm^{-3}$, $n_H=30\, cm^{-3}$, and $n_{H2}=150\, cm^{-3}$.

The spectral densities plotted in Figure 1 clearly demonstrate that Coulomb losses 
dominate up to relatively high energies, $\sim 1$ GeV in a SBN. In fact, this is so 
not only in the high density, high magnetic field environment of a SBN, but also 
across the disk of star-forming galaxies for which both the gas density, O(1) cm$^{-3}$, 
and magnetic field, $B<10\, \mu G$, are (correspondingly) lower than in a SBN, so that 
the ratio $n_{i}^{1/2}/B$ is roughly comparable to that in a SBN. However, the highest 
characteristic frequency ($\propto B$) at this critical value of $\gamma$ is obviously 
much higher in a SBN than its typical value across the disk of a star-forming galaxy.

\section{Synchrotron and Compton Spectra}

The standard formula for the synchrotron emissivity by an isotropically distributed 
population of electrons with PL spectral density $N_{1}\gamma^{-q_{pl}}$ 
is (Blumenthal \& Gould 1970)
\be
j_{pl}(\nu) = \frac { \sqrt{3}e^{3} B N_{1} } {4\pi m c^{2} } \int sin(\theta) d\Omega 
\int_{\gamma_1}^{\gamma_2} \gamma^{-q_{pl}} d\gamma \frac {\nu}{\nu_c} 
\int_{\nu/\nu_{c}}^{\infty} K_{5/3}(\xi)d\xi \, , 
\label{eq:synch_n_e_p}
\ee
where $e, \,m, \,c$ have their usual meaning, $B$ is the mean magnetic field strength 
across the emitting region, $\theta$ is the pitch angle,  $\nu_{c} = \nu_{0}\gamma^{2} 
sin(\theta)$ is the characteristic synchrotron frequency, and $\nu_{0} = 3 e B/ (4\pi m c)$ 
is the cyclotron frequency.
The last integrand is the modified Bessel function of the 2nd kind, with an integral 
representation
\be
K_{5/3}(\xi) = \int_{0}^{\infty} \exp^{-\xi cosh(t)} cosh(5t/3) dt \, . 
\ee
It is usually assumed that for all frequencies of interest $\nu_{c}(\gamma_1) \ll \nu \ll 
\nu_{c}(\gamma_2)\,$, \, so that the $\gamma$ integration can be well approximated by 
integration over the interval $[0, \infty]$. Doing so yields the standard formula
\be
j_{pl}(\nu)=   \frac {4\pi  e^{3}} {m c^{2}} a(q_{pl}) N_{1} B\biggl (\frac 
{\nu_{0}}{\nu}\biggr )^{(q_{pl}-1)/2} \, ,
\label{eq:synch_n_e_p_f}
\ee
where $a(q_{pl})$ is expressed in terms of a ratio of $\Gamma$ functions (eq. 4.60 of 
Blumenthal \& Gould 1970).  

For the steady-state electron density specified in eq. \ref{eq:n_e_ss}, the synchrotron 
emissivity is 
\be
j(\nu) = \frac {  \sqrt{3}e^{3} B  k_{i}} {2 m c^{2} (q_{i}-1)} 
\int_{0}^{\pi} sin(\theta) d\theta \int_{\gamma_1}^{\gamma_2} \frac{ \gamma^{-(q_{i}-1)}}
{b(\gamma)} d\gamma  \frac {\nu}{\nu_c} \int_{\nu/\nu_{c}}^{\infty} K_{5/3}(\xi)d\xi \, .
\label{eq:synch_n_e_s}
\ee
Using the above integral representation for $K_{5/3}(\xi)$, affecting the change of 
variable $x=\nu /[\nu_0 \gamma^{2} sin(\theta)]$, and extending the limits of 
the $\gamma$ integration to $[0, \infty]$, we obtain the following 3D integral
\be
j(\nu) = \frac {\sqrt{3} k_{i}} {4 (q_{i}-1)} 
\frac {e^{3} B} {m c^2}
\biggl (\frac {\nu_0}{\nu}\biggr )^{\frac {q_{i}} {2}}  
\int_{0}^{\pi} sin(\theta)^{\frac {q_{i}+4}{2}} d\theta
\int_{0}^{\infty} \frac { x^{q_{i}/2} dx } {b(\nu,\theta, x)}
\int_{0}^{\infty} e^{-x cosh(t)}\frac{cosh(5t/3)}{cosh(t)} dt \, ,
\label{eq:synch_n_e_s_2}
\ee
where 
\be
b(\nu,\theta, x) =  b_{0} (\nu_{0}/ \nu) sin(\theta) x 
+ b_{1} [(\nu_{0}/ \nu) sin (\theta) x ]^{1/2} +b_2 \, . 
\ee
(In a randomly oriented magnetic field the $\theta$ integral can be put in a closed form in terms 
of a combination of Whittaker functions [Crusius \& Schlickeiser 1986].)

To compare the above curved steady-state spectrum to that of a PL distribution, 
we determine the spectral index that fits the most relevant range of 
the measured radio spectrum, and select a relation between the normalization factors 
$N_{1}$ and $k_i$. Given that energetic electron spectra are usually determined from 
radio emission, the relative normalization of the two distributions can be based on the 
measured flux at some characteristic frequency $\nu_c$, $j_{pl}(\nu_c) = j(\nu_c)$. 
Below we compare results obtained with energy density normalization (specified 
in the previous section) to those obtained with the radio flux normalization.

Note that since the decay of charged pions (produced in proton-proton interactions) yields 
secondary electrons and positrons, which obviously contribute to the total synchrotron and 
Compton emission, the contributions of secondary electrons and positrons are essentially 
accounted for {\it approximately} by normalizing the electron spectral density to the 
measured (i.e., total emitted) radio flux. In our estimates of particle energy densities 
(in the next Section) this is quantified through the inclusion of the secondary-to-primary 
ratio, $\chi$.

Since the most relevant range of the measured radio spectrum is $\sim 1-10$ 
GHz, and given that typically galactic radio spectra in the inner disk region are well 
fit by $\sim 0.65$, the implied PL distribution is characterized by $q_{pl}=2.3$. Of 
particular interest is the theoretically favored value $q_{i}=2$, which for the most 
relevant frequency range is also in accord with a mean value of $\alpha \sim 0.65$. 
To assess the significance of employing the more realistic steady-state spectral 
distribution in the analysis of radio measurements, we consider the intense star-forming 
environment of a nuclear SB region in which the gas density and mean magnetic field are 
much higher than typical values across the galactic disk. We calculate $j_{pl}(\nu)$ with 
$q_{pl}=2.3$, and $j(\nu)$ with $q_{i}=2$, $B=100 \,\mu$G, $n_i =200\, cm^{-3}$, $n_H=30\, 
cm^{-3}$, and $n_{H2}=150\, cm^{-3}$, typical values in the nearby SBN of M82 \& NGC253 
(e.g., Persic \& Rephaeli 2014). The spectra are shown in Figure 2 (in arbitrary units); the PL spectrum 
is shown for the case when the two distributions are normalized to the same flux at 5 GHz, 
and for the case when the distributions are normalized to the same energy density. 

The different energy dependence of the loss processes results in a curved radio 
spectrum, with nearly a flat spectrum at very low frequencies to nearly a PL  
with an asymptotic index $q_{i}/2=1$ at high frequencies, corresponding to emission 
from high energy electrons whose losses are synchrotron-Compton dominated. As is 
evident from Figure 2, the overall impact of strong magnetic fields and high Coulomb 
losses on the synchrotron spectrum is a flattening that extends to relatively high 
frequencies well into the measured GHz range. 
Clearly, the synchrotron spectra shown in Figures 2 \& 3 do not include the effect 
free-free absorption which decreases the emergent emission at very low frequencies 
(Condon 1982). Absorption is not included here primarily because 
our treatment is focused on the impact of spectral flattening due to electron energy losses, 
rather than on detailed modeling (and parameter extraction) in specific SBNs. In actual 
spectral fitting to the observed spectrum of a SBN, free-free absorption has to be modeled 
and accounted for before the spectral flattening due to particle losses can be determined, 
as was done by, e.g., Yoast-Hull et al. (2013, 2014a, 2014b).

\begin{figure}[h]
\epsscale{0.7}
\centering
%\plotone{fig2_synch_sbn_2.3_2nor.eps}
\plotone{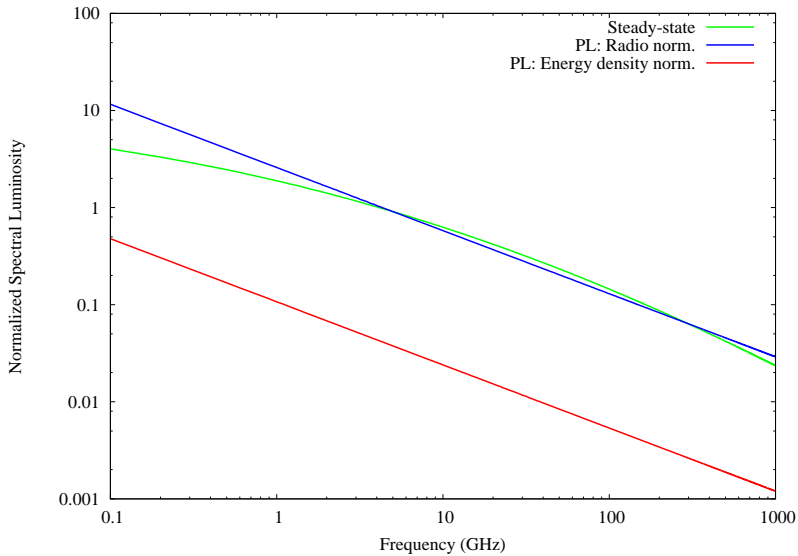}
\caption{Normalized synchrotron spectra for steady-state and PL density 
distributions in a SBN region. The steady-state spectrum is shown by the green line; PL spectra are 
shown by the red and blue lines, with the relative normalization set to either the same energy 
density or to the same 5 GHz emissivity as that of the steady-state population, respectively. 
Spectral indices are $q_{pl}=2.3$, and $q_{i}=2.0$, and $B=100 \, \mu$G.
\label{fig2}}
\end{figure}

Clearly, when the two distributions are normalized to the same energy density, 
the spectral emissivity of the PL is lower than that of the steady-state 
distribution. Since the steady-state spectrum was derived by explicitly accounting 
for electron losses at low energies, the spectral density can be extended down to 
the gas thermal energy, which is typically O($1$) keV. For our selected values of 
the PL index $q_{pl}=2.3$ and source injection spectrum with index $q_{i}=2$, the PL 
to steady-state energy density ratio reaches a value of $7.7$ at kinetic energy of 
511 keV ($\gamma_1=2$), and attains its maximal value $9.2$ already at kinetic energies 
that are still much higher than the thermal gas energy ($kT \sim 1$\,keV). The steeper the 
radio spectrum, the higher is the energy density ratio, which equals $17.4$ and $33.6$ for 
$\alpha = 0.7, \, 0.75$, i.e., for PL spectral densities with indices $2.4$ and $2.5$, 
respectively. 

\begin{figure}[h]
\epsscale{0.7}
\centering
%\plotone{fig3_synch_sbn_2.3.eps}
\plotone{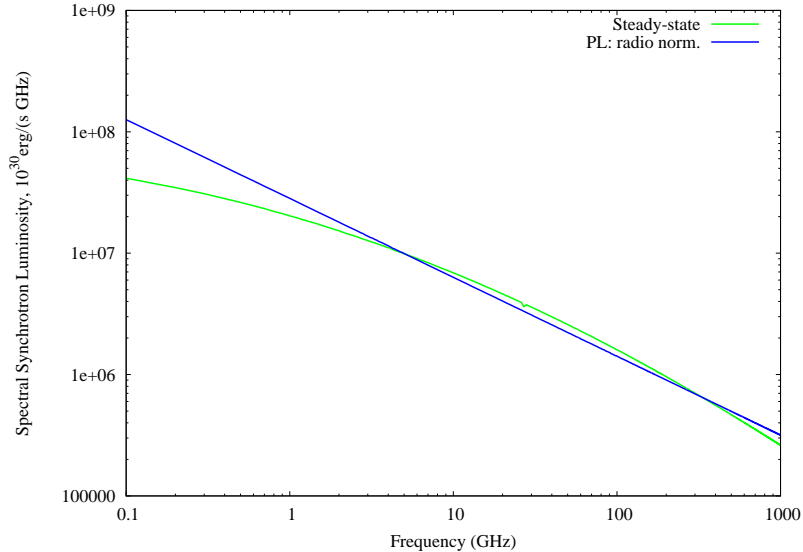}
\caption{Synchrotron spectral luminosity from electrons with steady-state (green line) 
and PL (blue line) distributions in a SBN. The distributions were normalized 
to the same spectral luminosity at 5 GHz, $L_{\nu}\simeq 2\times 10^{34}$ erg/(s Hz), 
with $q_{pl}=2.3$, $q_{i}=2.0$, and $B=100 \, \mu$G.
\label{fig3}}
\end{figure}

Since our treatment here is strictly spectral with no account taken of the spatial 
variation of the particle density across the galactic disk, extending the above 
description to the full disk is obviously unrealistic. However, our main objective 
here is to assess the impact of just replacing a PL distribution which, after all, 
is commonly assumed in characterizing particle spectra in galaxies, we carry 
out a similar calculation also for the full galactic disk. We realize, of course, 
that this is at best only a rough approximation of the more realistic description 
which necessarily involves a solution of a spectro-spatial kinetic equation. 
Repeating then the calculation with $B=5\, \mu$G, $n_i=1\, cm^{-3}$, and 
$\rho=4\rho_0$, which are typical mean values across the disk of a star-forming 
galaxy, and assuming a typical best-fit radio spectrum with $\alpha = 0.75$ (i.e., 
$q_{pl}=2.5$), we obtain a value of $31.0$ for the energy density ratio. 

The above large values of the energy density ratio reflect the fact that most of the 
energy density of the PL population is in lower energy electrons. Therefore, 
when the measured spectrum is fit by a PL, the density normalization is strongly 
weighted by the emissivity in the measured spectral range, and consequently the 
deduced energy density is much higher than that of the (curved) steady-state spectrum. 

Normalizing the electron spectral density by the measured radio spectrum allows 
a direct prediction of the electron spectral Compton luminosity. Of particular 
interest is a comparison of the predicted X-ray and $\gamma$-ray spectral luminosity 
of a PL distribution in a SBN to that of a steady-state distribution. Integrations 
of the spectral Compton emissivity of an electron scattering off a diluted Planckian 
radiation field (in the Thomson limit; see, e.g., eq. 2.42 in Blumenthal \& Gould 1970) 
over the electron PL and steady-state distributions, yield the spectra shown in Fig. 3. 
The distributions were normalized to the same spectral radio luminosity at 5 GHz, 
fiducially set to the measured flux from the SBN region of NGC253 at a distance of 3 Mpc.  

\begin{figure}[h]
\epsscale{0.7}
\centering
%\plotone{fig4_comp_sbn_2.3.eps}
\plotone{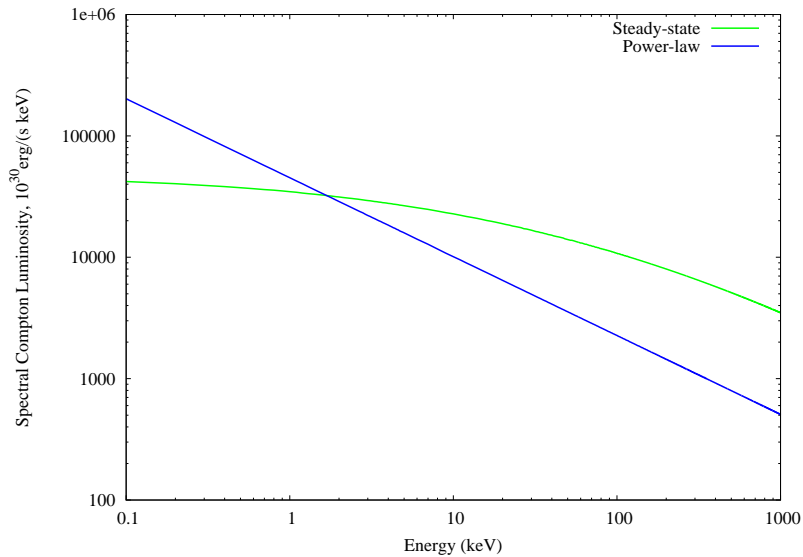}
\caption{Compton spectral luminosity of electrons with steady-state (green line) 
and PL (blue line) distributions in a SBN. The distributions were normalized 
to the same spectral radio luminosity at 5 GHz, , $L_{\nu}\simeq 2\times 10^{34}$ erg/(s Hz), 
with $q_{pl}=2.3$, $q_{i}=2.0$, $B=100 \, \mu$G, and diluted Planckian radiation 
field at a temperature $T=40$ K and energy density $\rho=40\rho_{0}$.
\label{fig4}}
\end{figure}

Comparison of the synchrotron and Compton spectra in Figures 3 \& 4 demonstrates an 
important consequence of the observationally-based normalization of a steady-state 
electron distribution: The Compton yield of this distribution is significantly higher 
than that of the PL distribution that is normalized to the same radio flux. As 
is obvious from the latter figure, the spectral luminosity in the hard X-ray 
($\epsilon > 10$ keV) and $\gamma$-ray is more than a factor $\sim 5$ higher 
than that of the PL distribution. This is due to the fact that the electron 
spectrum is flatter at (a mean) energy $\epsilon = (4/3)\gamma^{2}\epsilon_0$,  
for which the Compton boost of the incident IR photon energy, $\epsilon_0$, yields 
an outgoing photon energy $\epsilon >10$ keV.

\section{Particle and Magnetic Field Energy Densities from Synchrotron Emission}

In most cases of interest radio synchrotron measurements are analyzed for the 
purpose of determining energetic electron properties, such as density and 
energy density, and the mean strength of the magnetic field in the emitting 
region. As argued above, the common assumption of a PL distribution 
which is fit to the radio data can lead to very inaccurate values for these 
densities. Such estimates are also quite meaningless since in most cases the 
contribution of low energy particles is altogether ignored: For typical electron 
PL indices $2.3-2.5$, the fractional energy density at low energies (but still 
relativistic) below 1 GeV is $\sim 90\% - 97\%$, without even accounting for 
(sub-relativistic) supra-thermal electrons. Moreover, in many applications 
energy equipartition is assumed in order to determine the value of the mean 
magnetic field; for this purpose the proton energy density needs to be estimated. 
This is usually determined by adopting a theoretical relation for the 
proton-to-electron number density or energy density ratio. Therefore, overestimation 
of the electron energy density leads to overestimation also of the proton energy 
density and the field strength. 

Quantifying the impact of the more realistic estimate of the electron energy density 
on estimates of the latter quantities necessitates a revised formulation of the common 
assumption of energy equipartition between energetic particles and magnetic fields: 
Since a state of equipartition is reached after a sufficiently long period of tight 
coupling between these nonthermal quantities, the asymptotic relation between their 
respective energy densities should be based on their steady-state distributions. 
Self-consistent calculation of these distributions (e.g., Rephaeli \& Silk 1995, Lacki \& 
Beck 2013) requires accounting for all energy loss processes of both electrons and 
protons, and for a description of the distributions across the full disk, also of their 
propagation modes. The quantitative description of such a more physically based 
equipartition state is outside the scope of our work here. As a first step towards this 
goal we describe here an approximate procedure that is based on the use (also) of 
a steady-state spectral density for the protons.

%The relevant proton energy loss processes in a mostly ionized gas are electronic excitations 
%(including ionization) 
The relevant proton energy loss processes in ionized and neutral gas are Coulomb (i.e., 
electronic excitations and ionization), and pion production in proton-proton interactions. 
The respective loss rates, $b_{p, C}(\gamma)$ and $b_{\pi}(\gamma)$, can be expressed 
by the simplified formulae (adopted from Mannheim \& Schlickeiser 1994) 
\begin{eqnarray}
%b_{p, ee} ~ \simeq~ 4.75 \times 10^{-16} \, n_{i} \frac{ \beta^2 } { \eta^3+ \beta^3 }
% ~ {\rm s}^{-1}\, ,
b_{C} ~ \simeq~ 3.32 \times 10^{-16} \, n_{i}\beta^2  \biggl [ \frac{ 1}{ \eta^3+ \beta^3 }
~ + ~\frac {1.2 (n_{HI}+2 n_{H_2})}{n_{i}} ~ \frac{(1+0.0185\,{\rm ln} \beta) } {
\beta_{0}^3 + 2\beta^3}\biggr ]  ~ {\rm s}^{-1}\, ,
\label{eq:ion_loss}
\end{eqnarray}
%with the temperature-dependent factor (i.e., proportional to the proton velocity) 
%$\eta = 2.0\times 10^{-3}\,(T/10^4\,{\rm K})^{1/2}$, and
for $\beta \geq \beta_0=0.01$, and the temperature-dependent factor $\eta=0.002 (T/10^4 {\rm K})^{1/2}$.
The loss rate due to pion production is  
\begin{eqnarray}
b_{\pi} ~\simeq~ 5.85 \times 10^{-16} ~ n_{i}  ~ (\gamma-1)  ~ \Theta(\gamma 
\geq \gamma_{\pi})~ {\rm s}^{-1}\,
\label{eq:pion_loss}
\end{eqnarray}
where $\Theta(\gamma \geq \gamma_{\pi})$ is a step function , and $\gamma_{\pi}=2.3$ 
corresponds to the threshold kinetic energy for pion production, $1.22$\,GeV. At low energies 
electronic excitations is the only loss process, whereas at energies above $1.22$\,GeV losses are 
mostly due to pion production. 
An example for this spectral change is the drop in the local Galactic proton spectrum at low 
energies recently measured by the Voyager I spacecraft  (Stone \ea 2013); as argued by 
Schlickeiser \ea (2014) this is due to Coulomb losses.
 
The proton injection spectrum is assumed to be PL in momentum which, when transformed 
to $\gamma$, can be written as $k_{p,i} \gamma (\gamma^{2}-1)^{-(q_{p,i}+1)/2}$. 
\be
\biggl(\frac{dN_p}{dt}\biggr)_i=k_{p,i} \gamma (\gamma^{2}-1)^{-(q_{p,i}+1)/2}\, ,
\ee
which is valid also at low energies ($\gamma \simeq 1$). At steady-state the spectral density is 
simply
\be
N_{p}(\gamma)=\frac{k_{p,i}(\gamma^{2}-1)^{-(q_{p,i}-1)/2}}{(q_{p,i}-1)(b_{C}+b_{\pi}) }\, .
\label{eq:ss_n_p_s}
\ee
The resulting spectrum is $N_{p}(\gamma) \propto (\gamma^{2}-1)^{-(q_{p,i}-1)/2}$ at 
energies for which the loss rate is dominated by electronic excitations, steepening to 
$N_{p}(\gamma) \propto (\gamma-1)^{-(q_{p,i}+1)/2} (\gamma+1)^{-(q_{p,i}-1)/2}$ 
at energies (above 1.22\,GeV) for which the loss rate is dominated by pion production. 

The energetic proton density is usually related to that of the electrons by assuming 
a near equality in the rates energetic protons and electrons escape the acceleration 
region (e.g., Pohl 1993, Schlickeiser 2002). 
In the spirit of our approach here (whose objective is to estimate the impact of replacing 
a PL with a steady-state distribution), we adopt the charge neutrality assumption. 
In context of our approximate spectral (rather than spectro-spatial) steady-state 
treatment, the very different particle densities in the SBN and galactic disk would be 
explained as a consequence of the higher density of acceleration sites in the SBN, on 
the one hand, and the higher energy loss rates there on the other hand. 
If so, and equipartition is indeed attained, radio measurements then provide the 
observable needed to determine the proton energy density and mean strength of the 
magnetic field. Use of the more realistic particle steady-state distributions, and 
consideration of the full spectral range of particle energies, appreciably improve 
on the standard calculation which is commonly based on PL spectra with a relatively 
high lower energy cutoff. Specifically, we extend our numerical example representative 
of conditions in the SBN of M82 and NGC253 (specified above) by computing also the 
proton energy density. 

The proton source spectrum is a PL with nearly the same index as for the electrons, i.e. 
$q_{p,i} \simeq 2$, and the best-fit PL spectrum with index $q_{p,pl}=2.2$, a value deduced 
from $\gamma$-ray measurements (Persic \& Rephaeli 2014, and references therein). Repeating 
the calculations in the latter work (where all parameter values are specified) 
%
%\footnote{
%We adopted these fiducial parameter values: an effective SBN radius of $250$\,pc, a 
%gas density $n_i=200$ cm$^{-3}$, $n_{H2}=200$ cm$^{-3}$, a secondary-to-primary 
%electron ratio $\chi=1.6$, a distance of  $3$\,Mpc, a radio flux density $f(10\,GHz)=1$\,Jy, 
%and electron and proton spectral indices $q_{pl}= 2.3$, $q_p= 2.2$; for more details, see 
%Persic \& Rephaeli (2014).} 
%
but with the steady-state spectra obtained here, we compute 
$\rho_p \simeq 75$ eV cm$^{-3}$ and $B \simeq 56\, \mu$G, 
as compared with $\sim 330\, eV/cm^{-3}$ and $\sim 120\, \mu$G 
when assuming PL spectral density distributions. Even though approximate,  these values 
clearly show that the assumption of PL density spectra leads to significant overestimation 
of the electron energy density, and consequently also to overestimation of the proton and 
field energy densities (assuming equipartition).

\section{Discussion}

The results presented in the previous section make it clear that the assumption of 
a PL distribution for energetic electrons and protons leads to large overestimation 
of their energy densities in comparison with the values deduced when more realistic 
steady-state spectral density distributions are used. We derived these distributions 
(self-consistently) by accounting for all relevant energy losses. Our quantitative estimates 
of the factors by which the electron energy densities are overestimated when deduced 
from radio measurements are in the range $\sim 10 - 30$ for radio spectral indices in 
the range $0.65-0.75$, when calculated for typical conditions in a SBN and the disk 
of a star-forming galaxy. The mere fact that in some of the analyses based on a PL 
distribution a high value of $\gamma_1$ is assumed (supposedly circumventing the 
need for explicit accounting for the spectral flattening due to Coulomb losses) does 
not reduce the inadequacy of the (single-index) PL model: Clearly, considerations of 
energy equipartition are not valid when the particle energy density is calculated over 
only part of the energy range, particularly so when the relatively large contribution 
of the lower energy particles is ignored. 

Curved synchrotron spectra may also be a consequence of spatially inhomogeneous 
magnetic fields in the emitting region (e.g., Hardcastle 2013). Of course, particle 
spectra are intrinsically curved when the distribution is either time-varying (e.g., Torres 
\ea 2012) or before a  steady-state is reached. In our treatment here (which is essentially 
complementary to the aforementioned works) we calculate self-consistently the synchrotron 
spectrum emitted by electrons whose spectral density is curved due to 
%Coulomb, bremsstrahlung, synchrotron, and Compton 
all the relevant energy losses sustained while traversing a region with a uniformly 
distributed gas, magnetic, and radiation fields. The impact of this more realistic modeling 
of the particle spectrum is usually more important than that of spatial variation of the gas 
density and magnetic field, especially for estimation of the particle total energy density in 
high-density environments (such as a SBN).

Compton scattering of relativistic electrons by IR and optical radiation fields in 
SBGs (Rephaeli, Gruber \& Persic 1995, Wik \ea 2014a), and by the CMB in galaxy 
clusters (Rephaeli 1979, Wik \ea 2014b) is likely to contribute appreciably to their 
X-and-$\gamma$-ray spectra. As is clear from the spectra in Figure 4, estimating 
the level of hard X-ray emission by extrapolating the best-fit PL to the radio data 
can substantially under-predict the likelihood of its detection (and obviously affect 
motivation for its search). Moreover, the accuracy of joint analyses of radio and X-ray 
data can be improved substantially by using the properly normalized steady-state 
distribution for the electrons. 

As specified in the previous section, low energy protons lose energy effectively by 
ionization (in neutral media) and energy excitation (in ionized gas), in addition to 
catastrophic losses in proton-proton interactions. Here we accounted for these energy 
losses under typical conditions in a SBN, as has been done for conditions in galaxy 
clusters (Rephaeli \& Silk 1995) and SB galaxies (Lacki \& Beck 2013). The effectiveness 
of Coulomb interactions of low energy electrons and protons with gaseous media 
obviously result in much lower steady-state spectral densities of  low energy particles 
than what would be predicted from PL spectra. Overall, this implies that gas heating 
rates are correspondingly lower than estimated when assuming PL forms for the particle 
spectral density.

\acknowledgments

We are indebted to the referee for an expert and constructively critical review of the 
original version of this paper.  The referee's suggestions and the changes made in 
their implementation considerably improved the paper. Work at at UCSD was supported 
by a JCF grant. MP acknowledges the hospitality extended to him during a visit to UCSD.  
%
%\bs

\end{document}